Applied hybrid binary mixed logit to investigate pedestrian crossing safety at midblock and unsignalized intersection

Mohammad Ali Arman [a]*, Amir Rafe [b], Tobias Kretz [c]

a- Traffic Laboratory, School of Civil Engineering, Iran University of Science & Technology, P.O. Box: 16765-163, Narmak, Tehran, Iran.

ORCID: 0000-0001-5880-8713

Email: Mohammadali.Arman@gmail.com

Phone: +98 916 665 2746

* Corresponding author

b- Ramona Research Institute, No 25, Golha St. Sadat Abad, Tehran, Iran.

Email: Amir.Rafe@ramonari.com

c- PTV Group, Haid-und-Neu-Str. 15, 76131 Karlsruhe, Germany.

Email: Tobias.Kretz@ptvgroup.com

*Abstract*

Pedestrian's crossing from unsignalized locations at intersections or midblock locations is a risky decision that could lead to fatal accidents. Despite making a decision to accept a safe gap to cross the street is a personal choice, studying this phenomena and its affecting variables are crucial for accident analysis and traffic safety investigations. In this paper we used video-taped data to study pedestrian gap acceptance behavior at an unsignalized intersection and a midblock facility in Tehran, Iran. Multi-way ANOVA test indicates simultaneous effect of gender and child accompaniment have the highest effect on size of accepted gap and waiting time. In addition, number of rejected gap is mostly affected by the child accompaniment status. Investigation about critical gap shows in general, critical gap was bigger at unsignalized intersection and for women. We defined a latent variable named caution behavior based on some observable indicators and using structural equation modeling it was estimated and used as an input in a binary mixed logit model. This variable identified important in gap acceptance decision. Despite both structural equation and logit models are used in previous gap acceptance studies, but according to our best knowledge this is the first use of the hybrid mixed logit modeling approach in this field and we developed a new methodology to combine psychological and behavioral aspects of pedestrians' gap acceptance studies. Modeling approach shows pedestrian decision regarding acceptance or rejection of a gap to be highly influenced by the size of current gap, caution behavior and waiting time.

**Keywords:** *Pedestrian Gap Acceptance; Crossing Safety; Caution Behavior; Behavioral Hybrid Mixed Logit Model.*

# 1. Introduction

Pedestrians are the most vulnerable users in the transportation system with a higher level of accidents related to the weak transportation structures and safety in developing countries rather than developed countries[1]. According to the official statistics, 24.3 percent of accident casualties in Iran and 54.6 percent in Tehran are pedestrians [2,3]. This share is 14 percent in the USA [4], 14 percent in France, 17 percent in Germany, 23 percent in the UK [5] and 26 percent in China [4]. While more than half of pedestrian fatalities (51.1 percent) are above the age of 65, 39.6 percent of them are under 15 years old [2,3]. These rates are 19.7 percent and 6.2 percent for above 65 years old and under 15 years old in the USA [4]. Reports published by Police and Legal Medicine Organization indicate high speed of the vehicles and inability of pedestrian in exact prediction of gap size and consequently choosing an unsafe gap is the main cause of more than 57.1 percent of pedestrian accidents and cause of 67.2 percent of pedestrian fatal accidents. In addition in 30.9 percent of fatal accidents pedestrian used cell-phone in time of accident. Fig 1 summarised some statistics about pedestrian traffic accident in Iran.

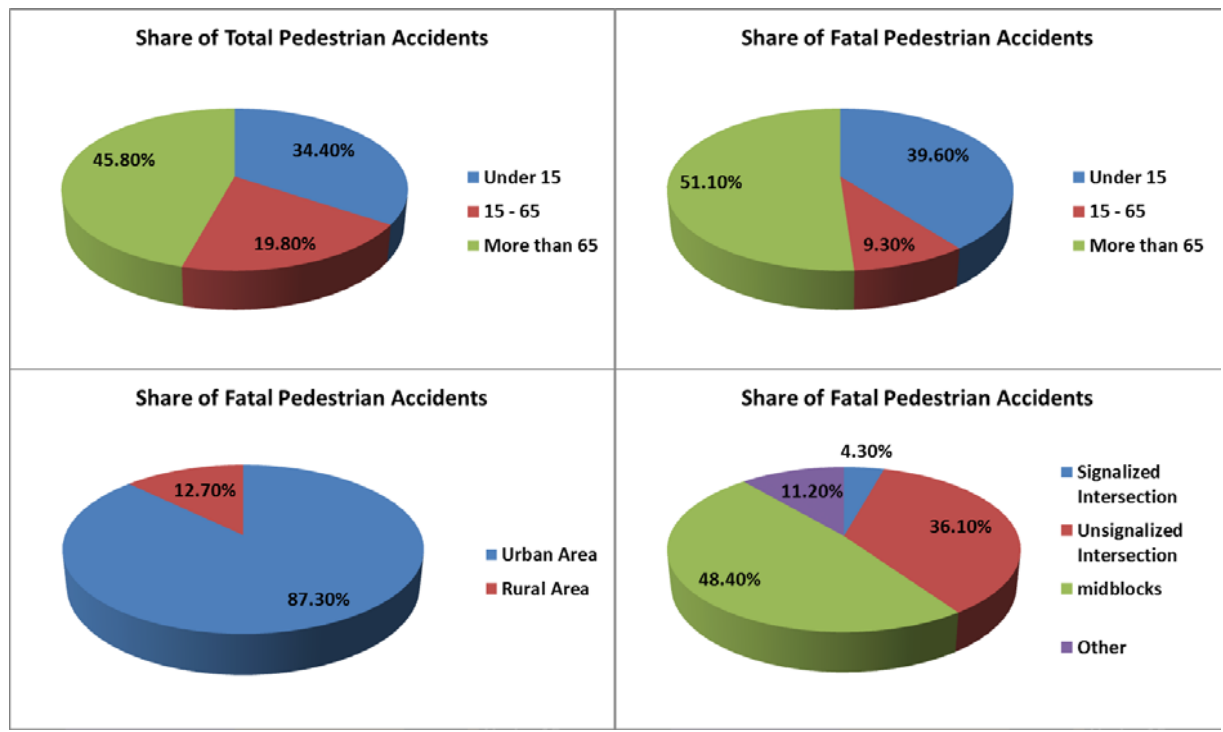

* Other includes parking lane/zone, bicycle lane, shoulder/roadside, sidewalk, driveway access, shared-use path/trail, non-traffic way area, and other.

**FIG 1. Pedestrian accidents and fatalities statistics in Iran**

In pedestrians' gap acceptance studies two subjects, safety and availability are important, however, because of the potential risk of conflict with vehicle flow the safety factor is more important. According to the literature, a gap is a time-interval/distance for a pedestrian between two successive vehicles. The pedestrian should make a decision about acceptance or rejection of the gap. In this paper gap acceptance studied based on time intervals.

---

[1] An in-depth comparison would also consider differences in modal split especially if the road traffic safety of children is concerned. In developed countries there has been a strong trend that parents drive their children by car to any destination while formerly they were walking. An example from Germany: In 1953 there were 2,090 children involved in accidents as passengers in a car, while 20,737 children were pedestrians in traffic accidents. Until 2011 the number of walking children in traffic accidents fell to 7,564, but the number of children hurt or killed as passengers in a car rose to 10,353 [1]. There was a maximum for both in 1970 (35,404 and 20,032) and since both numbers could be reduced, which indeed shows that there is success of the efforts for safety. Still, the comparison of both figures and their development over the years gives some insight into mobility behaviour dynamics.

Several factors could influence on the acceptance or rejection decision of a gap. Literature shows various variables were focused on by researchers that could be categorized below as:

- Pedestrian characteristics: the age and gender , walking speed [6-13], moving pedestrian in group [14-17,6,18-20,12,21], waiting time [14,15,17,18,8,9,11,20-22] and accompanying children or having bag/backpack [23,15,24,18,25,26].
- Vehicle properties such as the length, type and speed of the vehicle [27,14-17,7,19,28,29,9,30,22].
- Geometric design of the facilities [31,27,14-18,11,30,21].

Effects of the above factors could be investigated individually or combined to determine which kinds of categories have significant effect on pedestrians' decisions in crossing the streets. In different traffic conditions, pedestrians choose one of these types of crossing manner:

1. One-stage crossing: the pedestrian crosses when she/he finds the appropriate gap.
2. Two-stage crossing: the pedestrian crosses to the refuge island and considers the flow of vehicle movement at the nearside and then crosses by choosing a suitable gap.
3. Multi-stage crossing: the pedestrian crosses each lane of street when she/he finds an appropriate gap which is called rolling gap [31].

Among the mentioned cases, the first one happens in one-side streets typically. The second case is commonly in streets which end with an intersection/refuge island, and finally the last one was observed in high volume streets and crowded intersections.

Before 1980, studies of the pedestrian's safety crossing from intersection were carried out based on traffic flow features and researchers did not consider the analysis or simulation of the pedestrian's behaviours. Therefore, study of pedestrian gap acceptance does not have a significant background and only in the two past decades were a few investigations done in this scope.

Three methods used in investigating the pedestrian gap acceptance (individually or mixed) to analyse/estimate pedestrian gap acceptance and rejection. These methods are: (1) behaviour analysis, (2) modelling and (3) simulation.

Some studies used mathematical models such as linear regression and discrete choice to estimating gap size and crossing decision. Simulation is used to investigate the gap size in two approaches. The first approach is applying simulator devices to visualization of assumed facilities and locating pedestrian at studio and recording the gap acceptance data. This procedure is an ideal method to investigate on some factors like age and gender [10,32]. The second approach is utilizing the simulation software to evaluate the desired facilities [33,34].

This paper uses a statistical analysis and modelling approach in order to investigate the pedestrian gap acceptance behaviour in Tehran, Iran. An integrated hybrid model structure is presented to estimate the size of gaps through a linear regressing model, estimate caution behavior by a structural equation modeling technique and finally uses these two variables along with some other independent variables in a binary mixed logit model to estimate pedestrians' gap acceptance behavior. Based on our best knowledge, it is the first time that such an integrated model is used to study the pedestrians' gap acceptance behavior.

The rest of this paper is organized as follows: data collection method and site description are explained in section 2. Section 3 deals with the representative statistics and data analysis as well as comparison of our findings with other studies. Theoretical model is presented in section 4 and the results are discussed in section 5. Finally, section 6 will be the concluding remarks.

## 2. Data Collection and Site Description

We used video recordings to collect pedestrians' crossing behaviour data in two selected facilities in Sohrevardi St. in Tehran, Iran. The survey period includes three hours from 7:00 to 10:00. One of these facilities is an unsignalized intersection and another one is a midblock crosswalk. Two factors have been considered for facility selection: first, both facilities are selected in same street and as near as possible to each other to avoid social differences in study; second, there were enough high buildings near the facility to provide a suitable view for camera installation. On the whole, 1163 accepted gaps and 1435 rejected ones were observed and studied in unsignalized intersection respectively along with 1208 accepted gaps and 1812 rejected gaps in midblock crosswalk. Video recording was done using five cameras simultaneously in each of these locations as represented in Fig 2. As shown in this figure, both crossing sections are two-ways streets with two lanes in each sides. There was an island in the middle of the streets. Each lane is 3.15 meters in midblock crosswalk and 3.32 centimetres in unsignalized intersection crosswalk. In addition, the island has 65 centimetres width and 27.5 centimetres high in midblock crosswalk and 50 centimetres width and 30 centimetres high in unsignalized intersection crosswalk. Neither of crossing sections were occupied with Side Park during video recording. The standard pedestrian crossing width is 5 meters in Iran, and this standard observed in both locations. Both pedestrian crosswalks have marking without any traffic calming facilities before them and there were not any bus or taxi stations near the crossing areas.

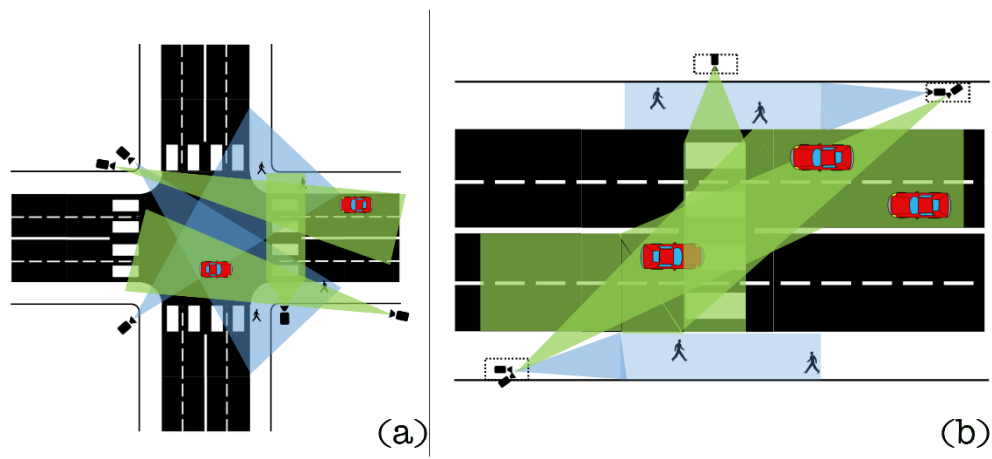

**FIG 2. Camera recording perspective**

Using video processing, required information in the three categories of geometrical features, pedestrians and vehicle characteristics were provided. These data analysed in section 3.

## 3. Representative Statistics and Data Analysis

As described previously, Data were collected using video recordings and processing. Based on these videos, pedestrian physical and behavioral attributes, conflict condition and vehicle and traffic flow features were collected.

## 3.1. Pedestrian Attributes

Pedestrian attributes are basically classified in two main groups: pedestrian physical characteristics and pedestrian behavioral attributes. Like most of the other studies, age and gender are the only two pedestrian physical characteristics. Pedestrian behavioral attributes include using cellphone, holding bags or briefcases, walking in groups and accompanying children. In unsignalized intersection, the behaviors of 623 pedestrians were observed and analyzed. This number was 607 in midblock crosswalk. Because pedestrians' age were estimated from the video[2], this property is divided into three discrete categories: young, middle-aged and old. Table 1 summarizes the gender and age distribution of pedestrians in two types of facilities studied in this paper.

---

[2] Because the high resolution video recording, categorization of the pedestrians based on appearance just in three categories was done with confidence. The age estimates has been evaluated an passable compromise in previous studies using designs similarity [35,27,8].

**TABLE 1. Number and percentage of observed pedestrians based on facility type, gender and age**

|  | Men | | | Women | | |
|---|---|---|---|---|---|---|
|  | *Young* | *Middle-aged* | *old* | *Young* | *Middle-aged* | *old* |
| **Unsignalized Intersection** | 375 (60.3%)(Total men) | | | 247 (39.7%)(Total women) | | |
|  | 110 (29.2%) | 197 (52.4%) | 68 (18.4) | 78 (31.4%) | 128 (51.9%) | 41 (16.7%) |
| **Midblock Crosswalk** | 371 (61.2%)(Total men) | | | 236 (38.8%)(Total women) | | |
|  | 112 (30.2%) | 196 (52.9%) | 63 (16.9%) | 73 (30.9%) | 123 (52.3%) | 40 (16.8%) |

Table 2 summarized the size of accepted gaps based on gender, age group and facility type. Women generally take larger gaps and are less risky. Similar results found by other researchers such as: [36,27,37]. Moreover, the accepted gaps in midblock crosswalk were smaller for both sexes.

Age is another factor that leads pedestrians to become more cautious. This factor was more effective in women. While changing age group from young to middle-aged and from middle-aged to old results in a 10.6% and 19.9% increase in accepted gap size between men in unsignalized intersection, these percentages are respectively 17.7% and 29.4% for women in this facility. This could be interpreted as age cautiousness effect. A similar pattern could be seen in midblock crosswalk: the above-mentioned percentages are 5.4% and 11% in midblock crosswalk facility for men and 9.4% and 21.1% for women. It is clear that aging effect has been more meaningful among women. In order to check the variables effect on waiting time and the size of accepted gap, multi-way ANOVA was used. The overall results are shown at the end of these discussions in Table 5.

**TABLE 2. Accepted Gap Size by Facility Type, Gender and Age in Seconds**

|  | *Average Size of Accepted Gap* | | | | | | *85th Percentile Size of Accepted Gap* | | | | | |
|---|---|---|---|---|---|---|---|---|---|---|---|---|
|  | *Men* | | | *Women* | | | *Men* | | | *Women* | | |
|  | *Young* | *Middle-aged* | *old* | *Young* | *Middle-aged* | *old* | *Young* | *Middle-aged* | *old* | *Young* | *Middle-aged* | *old* |
| **Unsignalized Intersection** | 9.44 (Men's average) | | | 11.09 (Men's 85th percentile) | | | 13.98 (Women's average) | | | 14.25 (Women's 85$^{th}$ percentile) | | |
|  | 8.41 | 9.37 | 11.43 | 8.89 | 11.03 | 15.4 | 12.62 | 13.89 | 16.43 | 12.78 | 14.21 | 16.98 |
| **Midblock** | 9.03 (Men's average) | | | 9.63 (Men's 85th percentile) | | | 12.98 (Women's average) | | | 13.48 (Women's 85$^{th}$ percentile) | | |
|  | 8.33 | 8.97 | 10.4 | 8.41 | 9.52 | 12.3 | 12.54 | 12.94 | 13.73 | 12.62 | 13.33 | 15.56 |

As mentioned previously, four variables are considered as pedestrians' behavior and their effects on the size of the accepted gap have been investigated in this paper. One of them that is not broadly studied (based on our best knowledge) is using cell-phone during crossing.

Generally, 41.2% of pedestrians in unsignalized intersection facility and 39.7% of them in midblock crosswalk facility used their cell-phone during crossing. In unsignalized intersection facility, 23.8% of pedestrians who confront with interaction with one or more vehicle used their cell-phone during crossing and this percentage was 20.4 in midblock crosswalk facility. This conduct is more prevalent among men (1.34 times more than women) and the youth act in this way approximately 2.1 times more than middle-aged people and 2.9 times more than the elderly. Based on traffic accident data in Tehran in 2013, on average 38.1% of all the pedestrians who were involved in a crash were using their cell-phone at the time of accident. Furthermore, 43.2% of fatal pedestrian accidents involved people who were using their cell-phone while crossing a cross street [7]. A similar effect of using cell-phones on pedestrian injuries and fatalities has been reported in other studies [38]. We found considerable reduction in pedestrians' awareness when they use their cell-phone during crossing and consequently they accept larger gaps, similarly other researchers reported reduced situation awareness and distracted attention among pedestrians using cell-phones [24,25,38-40]. As our results shown, in both genders and in all age groups, using cell-phones widened the accepted gaps. According to the results derived from Table 5, when the simultaneous effects of gender, age and using cell-phones during crossing are checked, F-value becomes larger compared to the effect of any of these variables individually or even in comparison with any of their paired compositions. Calculating the percentage change of accepted gap size in different age groups and genders that used their cell-phones during crossing shows that the most substantial increase is in age variation from the middle-aged to the elderly and also it is more remarkable in women compared to men. While the average and 85th percentage of the size of the accepted gap for men who used their cell-phones are 10.1 and 15.2 seconds respectively, these numbers are 12.0 and 15.8 for women. Moreover, young women normally need 11.6 seconds for crossing and older people need larger gaps, 16.0 seconds on average[3]. Similar to our findings, [41] conducted a study to examine the effect of using cell-phones during crossing in Seattle. The results showed 7.3% of pedestrians speaking with cell-phone during crossing and 11.2% of them text messaging, and it concluded that these pedestrians need a 1.87 second bigger gap (18%) to cross the street. Another research with results similar to ours is [26]. Their results show that accidents of men and women who use cell-phone are 52.9% and 58.7% respectively which indicates significant difference. The results show that 54.7% of pedestrians who have had accidents while using cell-phone are below the age of 31.

The next variable considered in this paper and which is found to be effective is accompanying a child during crossing. Usually it is expected that, children walking slower compared to adults[4]. Consequently it is reasonable to reduce the speed of the adults when they accompany a child and seek to find larger gaps. Moreover, it is suggested that adults become more cautious regarding accident risk when they accompany a child. From a gender perspective, this study found that in most cases children walked with a female (63.4%). Men's average and 85th percentile of accepted gap when they accompany a child are 11.7 and 16.4 seconds respectively; these numbers are 13.2 and 17.9 seconds for women. As could be concluded and presented in Table 5, through ANOVA results, considering gender and child accompaniment along with each other is more effective than any of these individual variables. Middle-aged pedestrians go along with a child more than other age groups and consist of 78.3% of all the observations, but the statistical evidence shows difference between sizes of accepted gap by different age groups of pedestrians who accompany a child. Similar to our results, O'Flaherty and Parkinson [42] observed women accompanying children had a 16% lower speed than single women.

Results of Table 5 indicate that if a pedestrian carries a bag or a briefcase, it does not have any effect on the size of accepted gap. These variables include bags that a single pedestrian could carry easily; other cases were too rare to be included in the study. Kadali and Vedagiri found a similar result in Hyderabad, India [18].

The last pedestrian behavioral variable that is investigated in this paper is walking in groups. Table 3 represents average and 85th percentile of accepted gap based on group size. Pedestrians may walk alone, in pairs, or in bigger groups when crossing the streets. Total observations consist of 66.5% of lone crossings, 16.8% of pairs, 7.3% of groups with three people and 9.4% of groups consisting of more people. As shown in table 3, for groups up to three persons, the accepted gap size becomes larger along with the group size, but it is not the case for groups consisting of more than three persons. A considerable reduction in the size of accepted gap is observed in such groups.

**TABLE 3. Accepted Gap Size Based on Group Size**

|  | *Walking Alone* | *Walking in Pairs* | *Walking in Groups with Three People* | *Walking in Groups with More than Three people* |
|---|---|---|---|---|
| ***Average Size of Accepted Gap*** *(second)* | 9.6 | 9.92 | 10.56 | 9.76 |
| ***85th Percentile Size of Accepted Gap*** *(second)* | 13.41 | 13.81 | 14.29 | 13.57 |

---

[3] In our database F-value for effect of age on size of accepted gap was 26.96 (see Table 5) and F-value for effect of pedestrians' speed on size of accepted gap was 31.87 and significant correlation (0.71) between pedestrians' age and speed was observed. So it is our opinion that in case of our study, size of accepted gap is more effected by age compared to speed and we could even suggest in our database speed may be a function of age.

[4] This assumption proved by previous study of authors: according to Shariat et al., (2012) studies in Tehran, an average walking speed of children is 0.93 meters per second and relatively 16% slower compared to that of the adults with average speed of 1.11 meters per second

We found some researches with results similar to our results and some in contrary with ours. The results of [20] indicate increase in the accepted gap by increasing the group size. [22] found if the pedestrian walks with another person, she/he will accept bigger gap compared to single crossing. Unlike the above results, [7] found by increasing the group size, the pedestrians accept smaller gaps.

An assumption that could be discussed and verified statistically is that the size of the accepted gap in group members depends on the group formation basis. Three main group formation basis could be found according to pedestrians' behaviors. Groups can be formed based on one or more of these reasons:
a) Pedestrians basically walk together in a group and cross the street in this form.
b) Long waiting time as a result of high vehicular speed or volume is imposed on pedestrians and during this time a group is formed that cross the street in the first appropriate gap with together.
c) A pedestrian or a group of them accept a suitable gap and cross the street, while another pedestrian joins them through a sudden acceleration of speed.

Whenever t-test did not show any statistical differences between the average sizes of the accepted gaps by group types (a) and (c), the average size of the accepted gap by group type (b) is approximately 11.3% smaller compared to other two groups. With respect to the group formation basis in group type (b), it seems that longer waiting times led pedestrians to become more aggressive and risk-taking, therefore accepting smaller gaps. Two-way ANOVA used to check the differences among the sizes of accepted gaps based on group sizes and group formation basis. The results showed that the size of accepted gap differed significantly across four groups sizes, $F(2,2368) = 11.59, p-value = 6.76e-4$ and it was also different among three group formation basis, $F(2,2368) = 6.375, p-value = 0.0116$.

In contrary with the above results, [7] and [20] concluded that when groups formed, pedestrians who endure longer waiting time accept bigger gaps.

As groups usually consist of different pedestrians with various behaviors, it is difficult to judge them based on behavioral features like using cell-phones or holding bags and briefcases. However, observations show that women cross more often in groups compared to men. Whereas 38.1% of women cross the street in groups, this percentage is 30.5 for men. In addition, compared to other age categories, the number of youth in groups is low, although most of the pedestrians that join a group based on pattern (c) are young. In contrast, elders cross in groups more than other age categories (similar results are reported by [17]) and they mainly belong to type (b). The larger the group size the slower the crossing and there is no exception for large groups as there was in case of the size of accepted gap. Among the remarkable basis for group formation, group type (c) is interesting: the numbers of those pedestrians with any slackening behavior in their crossing, like using cell-phones or accompanying a child, are relatively very low in these type of groups. Although men with more walking speed and accepting smaller gaps are often more agile than women, in case of joining and forming groups type (c) women are more numerous. This behavior could be interpreted in a sense that women are more cautious than men and prefer crossing safely in groups.

The number of 1163 accepted gaps for 623 pedestrians leads to a gap acceptance rate of 1.87 per person in unsignalized intersections. This rate equals 1.99 in midblock crosswalk. Therefore, it can be inferred that some pedestrians had to accept more than one gap to fully cross the street. Since both facilities are divided into two-way streets, pedestrians may be involved in conflicts with vehicles in slow-lane or passing-lane before (conflict on the left side) or after (conflict on the right side) reaching safety. In some cases, pedestrian did not wait for the street to be cleared and started crossing just after finding a proper gap in the nearest lane. In such kinds of behavior, known as rolling gap. A pedestrian may stop in the middle of the street after one successful gap acceptance and look for a suitable gap in the next lane; or she/he may cross the next lane immediately because it has been cleared after his/her first-line crossing. 18.7% and 26.5% of all the observed pedestrians exhibited the rolling gap behavior in unsignalized intersection and midblock crosswalk respectively. The number of men showing such behavior in both facilities was considerably more than women, with share of 82.8% and 84.6% from all of the rolling gap behaviors in unsignalized intersection and midblock crosswalk. Among the observed rolling gap behavior in unsignalized intersections and midblock crosswalks, in 29.7% and 31.1% of cases pedestrians had to find a proper gap in passing-lane after crossing slow-lane and in the rest of the cases passing-lane became clear during their first-line crossing.

We compared the size of the accepted gaps in slow-lane and passing-lane and found lane types exert a significant effect on the gap size, ($t(716) = 19.98, \text{p-value} < 2.2\text{e}-16$ for unsignalized intersection and $t(748) = 14.67, \text{p-value} < 2.2\text{e}-16$ for midblock crosswalk).

The direction of conflict was found not to be effective on the size of the accepted gap even while considering the facility type or people's gender; however, in an analogous pattern between two genders and two facilities, pedestrians wait less time to find gaps when they are involved in a right-side conflict compared to a left-side one.

Although gender and aging effected the waiting time of pedestrians, facility type was not found to be meaningful in pedestrians' waiting time. Based on the collected data, average and 85$^{th}$ percentage of waiting time for men are 9.3 and 26.5 seconds and statistically equal in both facilities (based on t-test). These numbers are 10.7 and 27.6 seconds for women. The results of multi-way ANOVA for checking different variables' effects on waiting time are presented in Table 5.

As expected, elders showed more patience and had the longest waiting time, while the youth were more impetuous compared to others. Again multi-way ANOVA was used to check the effect of different variables on waiting time and the results are presented in Table 5. As could be seen, both variables of gender and age are effective and their simultaneous effect is stronger (similar results are reported by [15]). Using cell-phones has the same effect on both the size of accepted gap and waiting time and cause to increase both of them. Pedestrian which used their cell-phones during gap acceptance crossing approximately wait 8.8% more than others. Considering the combinational effect of using cell-phones along with age and gender simultaneously is more operative than considering each of them separately. The average waiting time of young, middle-aged and elderly women is 10.2, 10.7 and 12.2 seconds and when women use their cell-phone these averages rise to 11.1, 11.9 and 13.5 seconds for these age groups. Similarly, accompanying a child leads to longer waiting times for both genders but not with an equal effect (11.3% increase for men and 20.2% increase for women). [15] found a similar result about the effect of child accompanying on waiting time in his research. As ANOVA test indicates, child accompaniment is more effective on waiting time compared to using cell-phones, whether they are evaluated solely or along with gender. Holding bags or briefcases again did not effect on the waiting time. Although grouping has no effect on waiting time, still the longer the waiting time, the larger the group size (similar results about the effect of waiting time on group size has been reported by [15]).

Whilst it is expected that the waiting time and number of rejected gaps being highly correlated for each pedestrian, Pearson correlation test between these two is 0.543. Again, ANOVA test was used to check the relationship between different variables and the number of rejected gaps by each pedestrian. In contrary to the size of the accepted gap and waiting time, the number of accepted gap is not affected by gender. Similar results were gained regarding facility type, using cell-phone and holding bags or briefcases. No statistically significant differences were observed between the youth and the middle-aged, but the older age group meaningfully rejected more gaps compared to other age groups. Finally, child accompaniment led to more gap rejection and it could be a result of more caution from adults for protecting children. As only age groups and child accompaniment were found to affect the number of rejected gaps, the simultaneous effect of these variables on the number of rejected gaps was not considered in ANOVA test. The results are presented in Table 5.

As mentioned before, five video cameras were used simultaneously to record pedestrian behaviors in sidewalks in both sides of the street and in the crosswalk. Moreover, the approaching vehicles were recorded from approximately 50 meters before reaching the crosswalk in both directions. Investigating pedestrians' speeds in sidewalks and crosswalks using t-test revealed that pedestrians increase their speed in the street where: $t(2445) = 17.72, \text{p-value} < 2.2\text{e}-16$. Fig 3 illustrated the cumulative distribution of speed plotted for men and women in sidewalk and crosswalk. The horizontal gray line indicates 85 percent.

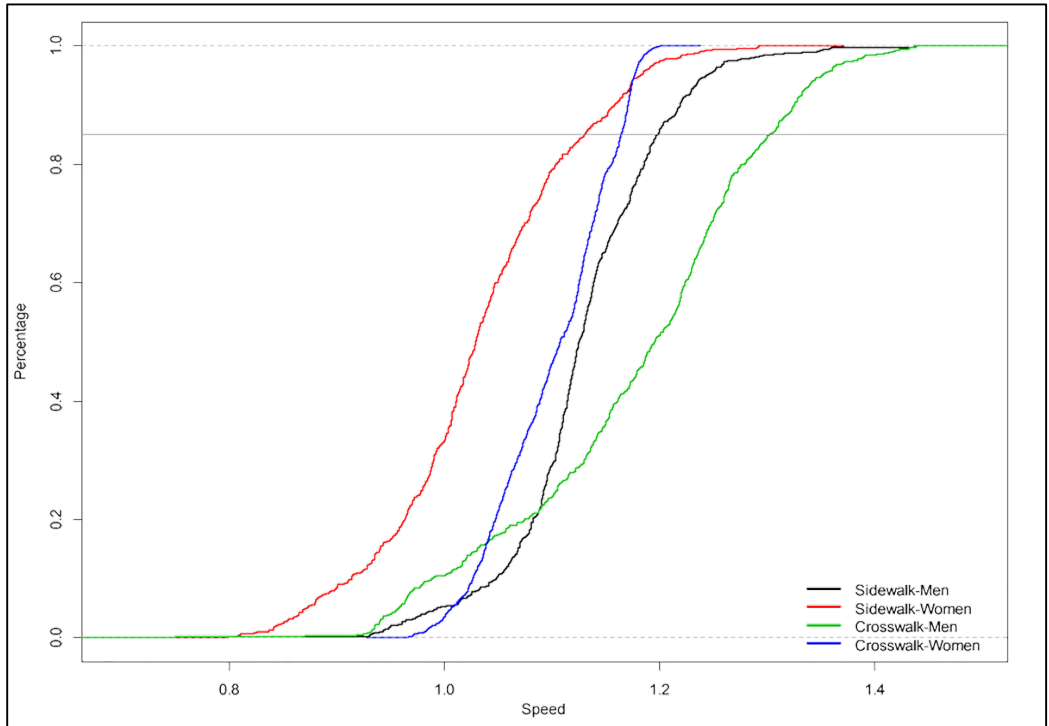

**FIG 3. Summary of Pedestrians Speed Study Results**

As described earlier, pedestrians increase their speed on the crosswalk, but this increase is not equal among both genders and age groups. Whereas men raise their speed relatively to 8%, 6.7% increase was seen among women. Here again the multi-way ANOVA was used to evaluate the effect of different variables on acceleration from sidewalk to crosswalk and the results are shown in Table 5. Gender was found to be the most effective factor on speed acceleration behavior (similar results are reported by [12]). Age, using cell-phone, child accompaniment and holding bags or briefcases are all effective factors in speed increase as well. Facility type is the only ineffective factor in this analysis and simultaneous consideration of gender with each of the other factors (except facility type) was found to be impressively more effectual. The most effective combination of factors is gender-child accompaniment. When pedestrians use their cell-phone, accompanying a child or holding bags or briefcases they increase their speed 0.8%, 1.5% and 1.3% lesser among men and 2.2%, 3.4% and 1.7% among women, compared to when they do not any of the three above mentioned behaviors.

### 3.2. Traffic Condition and Vehicles' Attributes

During three hours of data collection from 7 a.m. to 10 a.m., traffic condition was in peak and off-peak periods. The average of vehicle speed around the unsignalized intersection in peak and off-peak periods was 31.5 k/h and 38.6 k/h respectively and these numbers were 34.7 k/h and 39.8 k/h around the midblock location. Moreover, the traffic volume in peak and off-peak periods in signalized intersections was 1288 and 845 vehicles per hour and these numbers were 1235 and 796 in the midblock facility. All the observed vehicles in this study have been classified into four groups: Private Car, Taxi, Bus and Motorbike. Although the type of conflicting vehicle is not effective on the size of accepting gap, it affects decision-making regarding acceptance or rejection of a certain gap (Similar results are reported by [7,22]). More than 91% of conflicts with buses rejected. This number was even more for conflicts between a pedestrian and a motorbike, where in more than 98.5% of cases pedestrians reject gaps in conflict with motorbikes. High rate of rejection gaps in conflicts with buses could be related to their size, because pedestrians are more cautious in these cases. On the other hand, in conflicts with motorbikes, motorists usually have more speed and do some tricky maneuvers to pass faster and pedestrians prefer not to get involved in such hazardous situations. As a result of high speed of motorcycles, pedestrians accept bigger gaps (Similar results are reported by [7,18,14]).

In unsignalized intersections, in 22.7% and 75% of all the cases pedestrians conflict with a taxi and private car respectively. These rates were 18.2% and 77.3% in midblock crosswalk. In both facilities, on average 60% of conflicts with taxies and 52% of conflicts with private cars eventuate to gap rejection.

### 3.3. Critical gap

According to gap acceptance theory, there is a critical gap for each pedestrian [8]. Critical gap defined as the smallest time that pedestrians may accepts as a gap. [43]). It is expected that as waiting time become longer, pedestrians my become more hasty and accept smaller gaps. Pedestrians' critical gap is an important factor in safety analysis of pedestrians' crossing [31]. There are two main approaches to calculate the pedestrian critical gap. The first approach proposed by Greenshields in 1960. Based on this approach critical gap is the minimum acceptable gap which is undertaken by 50 percent of all of the pedestrians [44,45]. The second approach is proposed by Raff for vehicles and latter extended for pedestrians. According to Raff's approach, the critical gap is the intersection between curves of the accepted gaps and rejected gaps [11,27].

In this paper we used Raff's method to calculate the critical gap. Fig 4 shows the curve of the cumulative distribution function for accepted and rejected gaps in based on the gender at midblock crosswalk. Similar curves for unsignalized intersection illustrated in Fig 5.

According to equation (1), a Logistic function is used to fit curves on accepted and rejected gaps for men and women in midblock facility and unsignalized intersection. The value of model parameters as well as t-tests are reported in Table 4. In equation (1), $y$ is the cumulative percentage of acceptance or rejection of gapes, $x$ is the gap size, $a$, $b$ and $c$ are the Logistic function parameters.

$$y = \frac{a}{1 + e^{(b \times (x+c))}} \quad (1)$$

**TABLE 4. Values and t-tests of the parameters of fitted curves for accepted and rejected gaps with Logistic distribution**

| Models | a | | b | | C | |
|---|---|---|---|---|---|---|
| | Value | t-test | Value | t-test | Value | t-test |
| *Men's Accepted Gap in Midblock Facility* | 99.39 | 56.50 | -0.48 | -28.22 | -9.16 | -83.59 |
| *Men's Rejected Gap in Midblock Facility* | 99.87 | 88.92 | 1.09 | 42.75 | -4.97 | -166.66 |
| *Men's Accepted Gap in Unsignalized Intersection* | 95.38 | 33.20 | -0.52 | -14.41 | -9.72 | -52.45 |
| *Men's Rejected Gap in Unsignalized Intersection* | 126.11 | 86.19 | 0.77 | 82.46 | -3.73 | -106.53 |
| *Women's Accepted Gap in Midblock Facility* | 94.11 | 95.21 | -0.65 | -34.62 | -9.27 | -178.14 |
| *Women 's Rejected Gap in Midblock Facility* | 149.60 | 21.76 | 0.87 | 27.44 | -3.76 | -33.52 |
| *Women 's Accepted Gap in Unsignalized Intersection* | 94.60 | 84.39 | -0.75 | -30.43 | -11.39 | -228.48 |
| *Women 's Rejected Gap in Unsignalized Intersection* | 107.47 | 36.84 | 0.71 | 99.59 | -5.73 | -452.14 |

**TABLE 5. Multi-way ANOVA Test for Relation between Explanatory Variables and Size of Accepted Gap, Waiting Time, Number of Rejected Gaps and Percentage of Speed Increase in Crosswalk**

| | Size of Accepted Gap | | Waiting Time | | Number of Rejected Gaps | | Percentage of Speed Increase in Crosswalk | |
|---|---|---|---|---|---|---|---|---|
| | F Value | Pr (>F) | F Value | Pr (>F) | F Value | Pr (>F) | F Value | Pr(>F) |
| *Gender* | 42.24 | .75e-11 | 5.41 | 0.02 | 2.79 | 0.094 | 20.18 | 7.7e-6 |
| *Age* | 26.96 | 2.15e-7 | 3.93 | 0.047 | 5.15 | 0.023 | 17.2 | 3.59e-5 |
| *Facility Type* | 15.68 | 7.6e-5 | 0.25 | 0.617 | 0.03 | 0.863 | 2.87 | 0.09 |
| *Cell-phone* | 122.45 | 2e-16 | 9.79 | 1.79e-3 | 0.397 | 0.528 | 9.96 | 1.64e-3 |
| *Child Accompaniment* | 225.01 | 2e-16 | 12.21 | 4.92e-4 | 49.05 | 4.1e-12 | 11.13 | 8.74e-4 |
| *Bag* | 2.06 | 0.15 | 0.426 | 0.514 | 2.11 | 0.146 | 7.86 | 5.12e-3 |
| *Walking in Group* | 98.8 | 2e-16 | - | - | - | - | - | - |
| *Gender × Age* | 406.0 | 2e-16 | 26.97 | 2.42e-7 | - | - | 98.5 | 2e-16 |
| *Gender × Cell-phone* | 225.62 | 2e-16 | 16.05 | 6.55e-5 | - | - | 108.53 | 2e-16 |
| *Gender × Age × Cell-phone* | 408.65 | 2e-16 | 42.8 | 8.89e-11 | - | - | 49.68 | 3.03e-12 |
| *Gender × Child Accompaniment* | 515.39 | 2e-16 | 61.29 | 1.06e-14 | - | - | 31.75 | 2.17e-8 |
| *Test Degree of Freedom* | 5607 | | 1219 | | 1223 | | 1219 | |

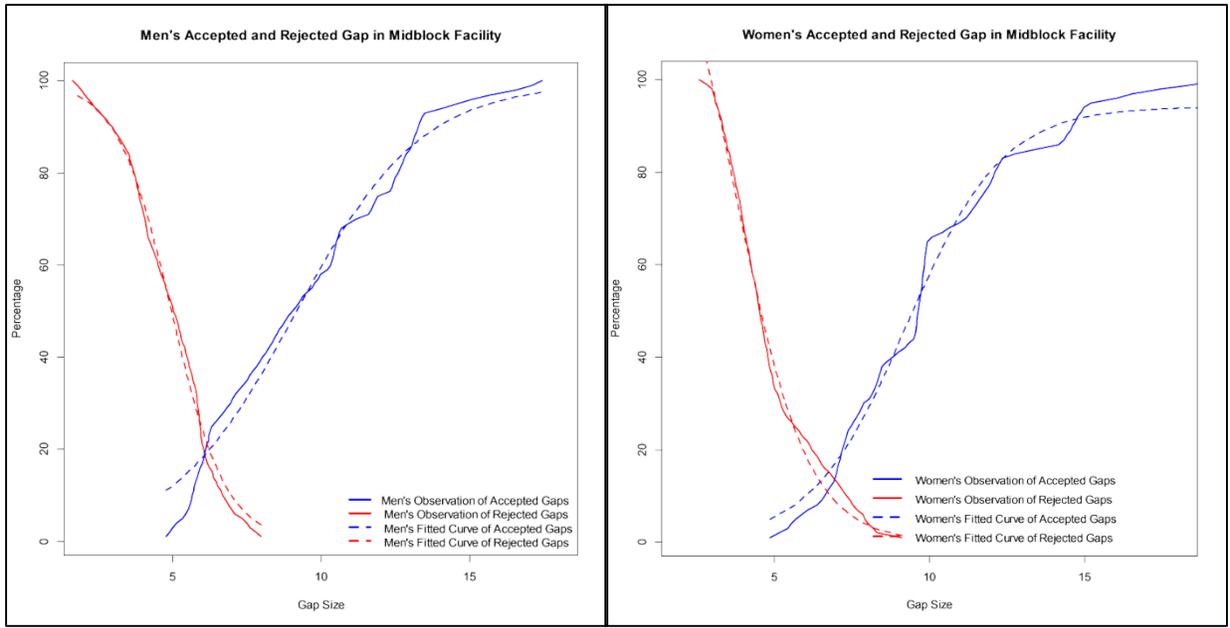

(A) Men's accepted and rejected gap  (B) Women's accepted and rejected gap

**FIG 4. Gender based cumulative distribution function of accepted/rejected gaps in midblock**

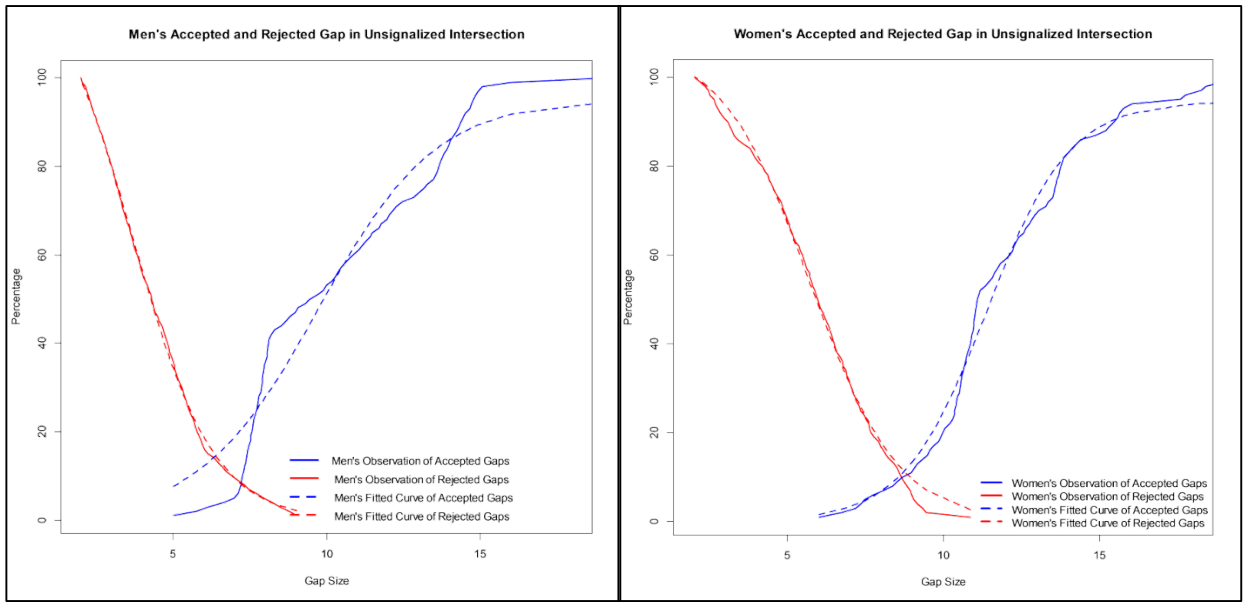

(A) Men's accepted and rejected gap  (B) Women's accepted and rejected gap

**FIG 5. Gender based cumulative distribution function of accepted/rejected gaps in unsignalized intersection**

Minimum accepted gap, maximum rejected gap, size of critical gap and its corresponding percentage for men and women in midblock facility and unsignalized intersection are summarized in Table 6.

**TABLE 6. Minimum accepted gap, maximum rejected gap, size of critical gap and its corresponding percentage based on gender and facility type**

|  | *Minimum Accepted Gap* | *Maximum Rejected Gap* | *Critical Gap Size* | *Critical Gap Percentage* |
|---|---|---|---|---|
| *Men's Gaps in Midblock Facility* | 2.68 Sec | 7.99 Sec | 6.08 Sec | 19.1% |
| *Women's Gaps in Midblock Facility* | 3.37 Sec | 9.10 Sec | 6.93 Sec | 13.7% |
| *Men's Gaps in Unsignalized Intersection* | 2.96 Sec | 9.03 Sec | 7.22 Sec | 8.7% |
| *Women's Gaps in Unsignalized Intersection* | 4.33 Sec | 10.87 Sec | 8.67 Sec | 9.8% |

## 4. Theoretical Model

In order to estimate the pedestrian gap acceptance behavior in our study, a Hybrid Binary Mixed Logit model was accommodated and used. Based on Fig 6 a regression and a structural equation model (SEM) component are used to estimate the gap size and pedestrian cautious behavior respectively. The outputs of these two components were used as inputs along with other independent variables in utility function of a mixed logit model.

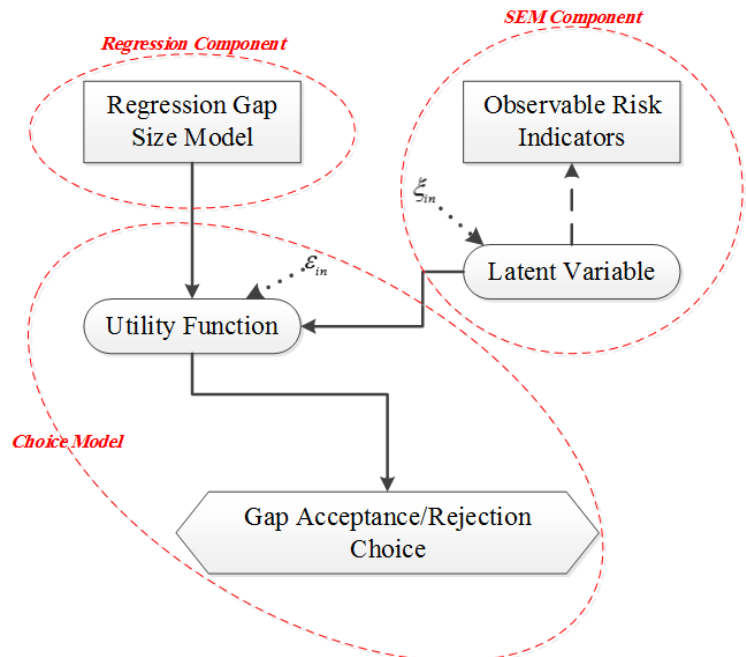

**FIG 6. Structure of hybrid binary mixed logit model**

The first part of this model structure is a linear regression model that estimates gap size based on independent variables.

The second part of our model is a Structural Equation Model. The idea of incorporating variables with structural equations into discrete choice models comes from the fact that combined effect of some of the variables may have stronger effect on pedestrians' gap acceptance decision compared to using them directly. In the other world we combine some of the variables to create an unobservable variable. We named this latent variable "cautious behavior". This latent variable constructed based on three observable indicators: gender, percentage of speed acceleration from sidewalk to crosswalk and the number of rejected gaps. Previous studies in both psychology [46] and traffic safety [47] show caution and risk perception to be different among genders. Consequently, gender was considered as an indicator in this latent variable. It was assumed that the higher the speed acceleration rate the greater the cautious behavior, because pedestrians were more likely to pass the street faster. As it was shown in section 3, child accompaniment requires more caution; therefore older people and pedestrians who accompany a child reject more gaps, and the number of the rejected gaps was involved as the third indicator. We believed pedestrians who rejected more gaps and wait for longer time would become more hurry and aggressive and consequently less cautious.

A binary mixed logit model is used to estimate pedestrian decision-making regarding the acceptance or rejection of gaps. As we could not judge about pedestrian utility or regret regarding the rejected gaps, a utility function estimated the accepted gaps and the utility of rejected ones was assumed to be zero. As the only difference between utilities is in important choice models, a comprehensive utility function for the accepted gaps could reflect the choice behavior perfectly. We used a Mixed Binary Logit model instead of a simple Binary Logit. The main motivation for the mixed logit model arises from allowing for random taste variation among decision makers. In the other word whereas the simple logit model estimate a fixed coefficient for every variables, this could be relaxed with estimation of a random variable with an assumed distribution as variables coefficient in mixed models. With this assumption for example being old is completely different with being young, but the effect of being old is not exactly same for all the old peoples. Any probability density function can be specified for the distribution of the coefficients in the population. The most widely used distribution is normal, mainly for its simplicity. The utility and probability function of choice model proposed in this paper are presented in equations 2 and 3 below:

$$U_{Acc-n} = \alpha\, GS + \beta \eta + \gamma x + C + \varepsilon \tag{2}$$

$$P_{Acc-n} = \int \left( \frac{e^{U_n^*}}{1 + e^{U_n^*}} \right) f(\mu) d\mu \tag{3}$$

Where:

$U_{Acc-n}$ is the utility of accepting a certain gap for pedestrian $n$,

$GS$ is the gap size that is the output of regression model,

$\alpha$ is corresponding coefficient of variable $GS$,

$\eta$ is the latent variable that reflects pedestrians' caution behavior and estimated by SEM technique,

$\beta$ is corresponding coefficient the latent variable,

$x$ is the vector of other independent variables in the utility function,

$\gamma$ is corresponding vector of coefficients of independent variables,

$C$ is the model constant and

$\varepsilon$ is the model error term.

In addition, $P_{Acc-n}$ represents the probability of accepting a gap; therefore $(1-P_{Acc-n})$ is the probability of rejection. $f(\mu)$ is the density function of random coefficients and $U_n^*$ is calculated with equation 4 and indicates pure difference between utility of accepting and rejecting a gap.

$$U_n^* = U_{Acc-n} - U_{Rej-n} \tag{4}$$

## 5. Results and Discussion

For testing the model, likelihood ratio test (755.8), Rho-square (0.361), Craig-Uhler (0.291) and Theil's tests (0.134) were used. All the tests revealed excellent prediction power of the proposed model. Theil's test has been used to control the overall model accuracy. The entire model has 17 variables and the estimated model was based on 3911 observations of accepted and rejected gaps. The ratio of observations to variables is 230.06 for his model. Tables 7 to 9 illustrate the model estimation results.

The first variable in regression model is conflict lane. This is a dummy variable: it takes one if the conflict lane is the passing-lane; otherwise, it is equals to zero. Due to a higher speed in passing-lane, a positive coefficient for this variable was expected which resulted in larger gaps in this lane.

With respect to section 3, conflict direction is not an effective variable on the size of the gap. To verify this claim, Pearson correlation test was used to check the correlation between this variable and the gap size. As shown in Table 7, this variable is not correlated with gap size, and consequently could be excluded from the regression model. The next variable is facility type. As shown in Table 7, this variable is meaningful in regression model. Facility type appears in this model as a dummy variable that is equal to one for unsignalized intersection and zero for midblock location. It means that, accepted gaps are larger in unsignalized intersection compared to midblock location. This finding confirmed with all of our previous analysis. The next variable in the regression model is length of the passing vehicle. Gap size is measured by the elapsed time between front of one the passing vehicle to that of the upcoming vehicle. Hence, the larger the size of the passing vehicle is a smaller remaining time for a pedestrian to complete a gap acceptance maneuver. Accordingly, the length of the passed vehicle appeared with a positive coefficient in the model and it has a remarkable correlation with the gap size.

[22], reported similar results about the effect of the length of the passing vehicle on size of the accepted gap. [7] reported similar results. They conclude the larger the size of the passing vehicle is a smaller probability of accepting gap.

The next variable in the regression model is speed of the upcoming vehicle. The higher speed of the upcoming vehicle leaves smaller crossing opportunity for pedestrians and they need bigger gaps to pass. This confirmed with a big positive coefficient for this variable in the regression model. Many other studies reported similar findings [9,7,19,14], however, the results of [16] have not shown any significant effect of vehicles speed on the size of the accepted gap. In addition, Lobjois and Cavallo in an experimental study [28] showed that when the pedestrians (all age groups) have time limit for crossing, they accept shorter gaps even at high speeds. Finally, although model constant is large, its t-test shows that it is not very meaningful in regression model compared to independent variable.

**TABLE 7. Model estimation results for regression model**

| Variables | Coefficient | t-test | DV-IV Correlation* |
|---|---|---|---|
| *Conflict Lane(Dummy)* | 0.84 | 8.47 | 0.602 |
| *Conflict Direction (Dummy)* | Not involved in the Model | | 0.083 |
| *Facility Type (Dummy)* | 0.72 | 4.48 | 0.489 |
| *Length of the Passing Vehicle* | 1.02 | 6.19 | 0.591 |
| *Speed of the Upcoming Vehicle* | 1.27 | 5.69 | 0.598 |
| *Model Constant* | -8.68 | -2.12 | - |

*DV-IV Correlation\*: Correlation between dependent variable (Gap Size) and each of independent variables.*

SEM was scaled to one for gender and loading factor estimated for the two other variables. Gender comes into the SEM as a binary variable. It is equal to one for women and zero for men. Consequently a loading factor equal to one in the model could be interpreted as more caution behavior of women compared to men. Between two other variables, namely "percentage of speed increase from sidewalk to crosswalk" and "number of rejected gaps", the first one is more important based on the loading factor. But as Table 8 shows, both variables have negative loading factor. It means pedestrians that increase their speed more compared to other ones are less cautious and pedestrians that rejected more gaps became more aggressive and their caution level decreased compared to other ones.

**TABLE 8. Model estimation results for SEM model**

| Variables | Loading Factor | t-test |
|---|---|---|
| *Gender* | 1 | - |
| *Percentage of Speed Increase From Sidewalk to Crosswalk* | -0.84 | -7.88 |
| *Number of Rejected Gaps* | -0.61 | -4.45 |

The first variable in the choice model is age. Although age is continuous variable, age of the pedestrians was recorded discretely, and accordingly this variable was presented in a dummy format in the model. Age group "Young" scaled to zero to prevent correlation error among three age group dummy variables in the model. Random coefficient based on normal distribution was estimated for age dummies which indicated taste variation among each age group. According to the results, negative coefficient of both "middle-age" and "old" groups compared to zero coefficient for "young" pedestrians indicate increasing of the age of the pedestrians leads to accepting larger gaps. As Table 9 shows variance of the coefficient for "middle-age" pedestrians is bigger compared to "old" pedestrians and it could interpreted as taste variation among them compared to elderlies and consequently larger variety of gap sizes that accepted by this group. Similar results found by other researchers such as [10,27,20,18]. T-test indicated that age is statistically meaningful in the model.

As described in section 3, using cell-phones has adverse effects on the size of the accepted gaps. Considering normal distribution assumed for this variable and big value that estimated for variance of the coefficient, the effect of using cell-phone on the size of the accepted gap is highly diverse among pedestrians. It could be a result of different effects of using cell-phone among genders and age groups. This variable was shown to be very significant by the t-test.

One of the most effective variables in this model is accompanying a child. This variable appeared with strong negative coefficient and the smaller variance (in the entire model) with normal distribution.

With respect to section 3, holding bags or briefcases is not an effective variable on the size of the accepted gap. In Table 9, a star mark is used to indicate that this variable is not significant in the model even in 85% of confidence level.

A random coefficient was estimated for the gap size based on the lognormal distribution. Because everyone prefers larger gaps, it is not reasonable to estimate a negative coefficient for this variable; consequently lognormal distribution is suitable. Variance of the coefficient for this variable is very small and it means pedestrians' taste variation is very limited regarding to the gap size.

The next variable in the model is caution behavior. Positive and big value of the estimated coefficient for this variable indicates caution is a determinant factor in pedestrian crossing behavior. In addition, relatively big value for variance of this variable shows pedestrians' behavior is not completely rational and same caution behavior could not expected from all the pedestrians.

Longer waiting time leads to more aggression and risk-taking behavior, so the pedestrians with longer waiting time are further encouraged to accept even the small gaps. To consider this suggestion, a random coefficient based on negative lognormal distribution was estimated for this variable. Compared to mean of the coefficient, the estimated variance is relatively big and it means waiting time and its aggression outcome has not a similar effect on gap acceptance behavior of pedestrians. Similar results about the effect of waiting time on pedestrians' gap acceptance reported in other studies [14,22,48].

As described in section 3, increase in group size leads to accepting larger gaps but it is not the case for groups larger than three persons. So it seems that taste variation should be considerable in this case. A random coefficient based on normal distribution was estimated to reflect this fact. Big value of estimated variance for this variable confirm this claim.

Finally, the model constant estimated fixed because a random constant pose heteroscedasticity in utility of choice alternative and this situation is not addressed in this model.

**TABLE 9. Model estimation results for mixed binary logit model**

| Variables | Coefficient Mean | t-test | Coefficient Variance | t-test | Selected Distribution Function |
|---|---|---|---|---|---|
| *Age* | | | | | |
| *Young (Dummy)* | 0 | - | - | - | - |
| *Middle-age (Dummy)* | -1.41 | -8.56 | 0.21 | 3.85 | *Normal* |
| *Old (Dummy)* | -3.02 | -9.64 | 0.06 | 3.91 | *Normal* |
| *Using Cell-phone (Dummy)* | -3.57 | -6.19 | 0.44 | 3.94 | *Normal* |
| *Child accompaniment (Dummy)* | -4.01 | -12.69 | 0.13 | 5.17 | *Normal* |
| *Holding Bag (Dummy)* | -0.39* | -1.72 | 1.17e-5* | 1.55 | *Normal* |
| *Gap Size* | 2.37 | 19.75 | 0.09 | 4.69 | *Lognormal* |
| *Caution Behavior* | 2.02 | 7.19 | 0.35 | 16.75 | *Normal* |
| *Waiting Time* | 1.88 | 8.43 | 0.33 | 2.94 | *Negative Lognormal* |
| *Group Size* | 0.89 | 5.12 | 0.85 | 7.16 | *Normal* |
| *Model Constant* | 2.54 | 10.09 | - | - | - |

*\* Variable is not meaningful in 85 percent confidence level.*

## 6. Conclusion

This paper tried to report an experimental analysis of pedestrians' crossing safety in unsignalized intersections and midblock crosswalks.

The statistical analysis of pedestrian decisions regarding acceptance or rejection of a gap and waiting times revealed that gender, using cell-phones and child accompaniment are extremely effective on pedestrians' behaviour.

A linear regression model was used to predict the size of the gap that pedestrians have to evaluate based on their norms. As expected, the gap size was determined by traffic conditions and then evaluated through pedestrians' behavioural norms.

Structural equation modelling technique was used to estimate a latent variable. This variable is considered in this paper as an index of pedestrians' caution behaviour and it was demonstrated that gender and percentage of speed increase in the street are the most dominant variables in this behaviour.

The last part of our study was a binary mixed logit model that used output of SEM as an input to consider behavioral factors into account. According to our best knowledge it is the first hybrid use of SEM and discrete choice models in the context of pedestrian safety analysis. The binary mixed logit model revealed that although the gap size is very important by itself, behavioural factors of pedestrians significantly affect their judgment to accept or reject a gap. Caution behaviour, child accompaniment and using cell-phones were found to be the most important behavioural factors in our study. As expected, the longer the waiting time, is the more aggressive pedestrians' behaviour. Consequently this leads to acceptance of smaller gaps. In addition age identified as an important variable in our model. As explained in section 1, Tehran's streets are specially dangerous for children and elderly pedestrians and wrong estimation of gap size by pedestrians is one of the most dominate cause in fatal pedestrian accidents that is more common among these two age groups. Despite there are many studies and empirical proceedings around the world to increase children's and elderly's safety in urban traffic [49,50], there is not any example in Iran and it could be a direction for further researches.